\documentclass[pdflatex,sn-mathphys-num]{sn-jnl}

\usepackage{graphicx}%
\usepackage{multirow}%
\usepackage{amsmath,amssymb,amsfonts}%
\usepackage{amsthm}%
\usepackage{mathrsfs}%
\usepackage[title]{appendix}%
\usepackage{xcolor}%
\usepackage{textcomp}%
\usepackage{manyfoot}%
\usepackage{booktabs}%
\usepackage{algorithm}%
\usepackage{algorithmicx}%
\usepackage{algpseudocode}%
\usepackage{listings}%
\usepackage{hyperref}
\usepackage{color}
\usepackage{setspace}
\usepackage{verbatim}
\usepackage{upquote}

\usepackage{colortbl}
\definecolor{gray_1}{RGB}{142,152,178}
\definecolor{mygray}{gray}{0.95}

\newcommand{\smalltt}[1]{{\small {\tt #1}}}
\newcommand{\mytilde}{\raisebox{0.4ex}{\texttildelow}}
\newenvironment{packed_itemize}{
\vspace*{-0.3em}
\begin{itemize}
\setlength{\partopsep}{0pt}
\setlength{\itemsep}{1pt}
\setlength{\parskip}{0pt}
\setlength{\parsep}{0pt}
}{\end{itemize}}

\AddToHook{cmd/ttfamily/after}{\frenchspacing}
\theoremstyle{thmstyleone}%
\theoremstyle{thmstyletwo}%

\theoremstyle{thmstylethree}%

\begin{document}

\title[TPTP World Infrastructure for Non-classical Logics]{\center TPTP World Infrastructure for\\Non-classical Logics}

\author*[1]{\fnm{Alexander} \sur{Steen}}\email{alexander.steen@uni-greifswald.de}

\author[2]{\fnm{Geoff} \sur{Sutcliffe}}\email{geoff@cs.miami.edu}

\affil*[1]{\orgdiv{Institute of Mathematics and Computer Science}, \orgname{University of Greifswald}, \orgaddress{\street{Walther-Rathenau-Str. 47}, \city{Greifswald}, \postcode{17489}, \country{Germany}}}

\affil[2]{\orgdiv{Department of Computer Science}, \orgname{University of Miami}, \orgaddress{\street{1365 Memorial Drive}, \city{Miami}, \postcode{33146}, \state{FL}, \country{USA}}}

\abstract{
The TPTP World is the well established infrastructure that supports research, development, and 
deployment of Automated Theorem Proving (ATP) systems.
The TPTP World supports a range of classical logics, and since release v9.0.0 has supported
non-classical logics.
This paper provides a self-contained comprehensive overview of the TPTP World infrastructure for 
ATP in non-classical logics:
the non-classical language extension, problems and solutions, and tool support. 
A detailed description of use of the infrastructure for quantified normal multi-modal logic is 
given.
}


\maketitle

%--------------------------------------------------------------------------------------------------
\section{Introduction}
\label{Introduction}

The TPTP World~\cite{Sut17} is the well established infrastructure that supports research, 
development, and deployment of Automated Theorem Proving (ATP) systems.
The TPTP World includes the TPTP problem library~\cite{Sut17}, 
the TSTP solution library~\cite{Sut10}, 
standards for writing ATP problems and reporting ATP solutions~\cite{SS+06,Sut08-KEAPPA}, 
tools and services for processing ATP problems and solutions~\cite{Sut10}, 
and it supports the CADE ATP System Competition (CASC)~\cite{Sut16}.
Various parts of the TPTP World have been used in a range of applications, in both academia and 
industry.
The web page of the TPTP World provides access to all components.\footnote{%
\href{https://www.tptp.org}{\tt www.tptp.org}.}

The TPTP World supports a range of classical logics, and since release v9.0.0 has also supported
non-classical logics. 
Non-classical logics are required for many real-world applications of ATP~\cite{SB24}, including
artificial intelligence (e.g., knowledge representation~\cite{GG+11}, planning~\cite{LAR20}, 
multi-agent systems~\cite{CLS23}), philosophy (e.g., formal ethics~\cite{BPT20}, 
metaphysics~\cite{BW16}), natural language semantics (e.g., generalized quantifiers~\cite{vBe87}, 
modalities~\cite{Kra77}), and computer science (e.g., software and hardware 
verification~\cite{Bry91}).
There has been a disconnect between classical and non-classical logics in the practical 
development and handling of automated reasoning technology, with classical logics receiving 
greater attention.
This is unfortunate because there exist ATP systems for non-classical logics, but their usage, 
interoperability, and incorporation within larger contexts was hampered by their heterogeneous 
input formats and non-uniform modes of result reporting.
The TPTP World now provides a homogeneous integration of ATP for classical and non-classical 
logics.

The project of extending of the TPTP infrastructure to support non-classical logics started
around 2015, with first experiments~\cite{WSB16} on representing modal logics~\cite{BBW06} and 
hybrid logics~\cite{AtC07}.
Since then the infrastructure has evolved
% \footnote{Thanks, to a large extent, to valuable community 
% feedback throughout the development process, e.g. during TPTP Tea Parties and direct email 
% communication.} 
to cover more non-classical logics, to reflect the natural representation of the logical 
expressions, to adhere to the standards and conventions of the TPTP World, and to be uniform 
across the first-order and higher-order variants of the TPTP language.
Based on previous work~\cite{WSB16,GSB17,GS18,Ste22,SF+22,SS24}, this paper provides a 
self-contained comprehensive overview of the TPTP World infrastructure for ATP in non-classical 
logics~\cite{Pri08,Gob01}.

While the non-classical TPTP syntax is designed to be able to represent many different logics, the 
non-classical logics standardized by the TPTP so far are quantified normal multi-modal 
logic~\cite{FM23} with specialized representations for alethic~\cite{Sch06-CPL}, 
deontic~\cite{Hil71}, epistemic~\cite{vDH15}, and doxastic~\cite{Hin62} contexts, and 
instant-based temporal logics~\cite{GR22}.
In this paper infrastructure is exemplified in quantified normal multi-modal logic~\cite{BBW06}.

\paragraph{Related work.}
A well-established format for propositional modal logic problems is provided by the Logics 
Workbench (LWB) benchmark~\cite{BHS00}. 
While the formula syntax was originally developed within the scope of a single benchmark set, it 
is now widely accepted by many propositional modal logic ATP systems. 
The LWB syntax focuses on only the propositional fragment of modal logic, and does not provide 
syntactic means for controlling or documenting the reference logic in which a problem is stated.
A representation format for quantified modal logics was developed and applied in the 
QMLTP project~\cite{RO12}. 
It contributed significantly to the practical application of first- and higher-order theorem 
provers in that context~\cite{BOR12}.
The syntax of the QMLTP focuses on unary modal connectives, and does not support the 
much broader range of non-classical logics that are the target of the TPTP World, e.g., logics 
that require support for arbitrary $n$-ary connectives that might take parameters, or 
generalizations of modal connectives indexed by, e.g., terms or formulae. 
Furthermore, the QMLTP covers only a small subset of relevant modal logics.
The DFG syntax~\cite{HKW96}, a format for problem and proof interchange developed in the 
DFG Schwerpunktprogramm Deduktion, contains a meta-information tag called {\tt logic} 
that can be used to specify ``non-standard quantifiers or operators'' in informal natural language.
This has had some limited use for modal logic~\cite{HS97-IJCAI,HS02}.
The Knowledge Interchange Format (KIF)~\cite{GF92} is a comprehensive format for knowledge 
representation, including numbers, lists, sets, and non-monotonic rules. 
KIF could be considered a language for non-classical logic. 
However, KIF is based on a first-order logic and comes with a fixed semantics. 
It is not flexible enough to capture different non-classical logics.
Common Logic (CL) is an ISO standard~\cite{II18} for the representation of logical information,
with several dialects and a common general XML-based syntax. 
While CL can express both first-order and higher-order concepts, it also comes with a fixed 
semantics.

\paragraph{Paper structure.}
Section~\ref{Preliminaries} provides a review of the classical TPTP languages, their underlying 
logics, and of different variants of first-order modal logic.
Section~\ref{ClassicalTPTPLanguages} reviews the classical TPTP languages, leading to
Section~\ref{NonClassicalTPTPLanguages} that introduces the non-classical TPTP languages.
Section~\ref{QMLinNTF} exemplifies the framework for quantified normal multi-modal logic.
Section~\ref{TPTP} gives an overview of the concrete integration of non-classical logic
into the TPTP World.
Section~\ref{Conclusion} summarises what has been achieved, and discusses plans for further
development.

%--------------------------------------------------------------------------------------------------
\section{Preliminaries}
\label{Preliminaries}

The underlying logics of the well-established classical TPTP languages are untyped and typed 
first-order logic (FOL)~\cite{MA22}, typed first-order logic with first-class Booleans 
(FOOL)~\cite{KKV15}, and higher-order logic (HOL) with Henkin semantics~\cite{Hen50,And72,BBK04}. 
It is assumed that the reader is familiar with these underlying logics, so that no review is
necessary.

First-order modal logic is a family of logic formalisms extending classical first-order
logic with the unary modal connectives $\Box$ and $\Diamond$. 
First-order multi-modal logic (FOMML) further generalizes the structure with multiple modalities,
$\Box_i$ and $\Diamond_i$, $i \in I$, where $I$ is some index set of constants~\cite{Bal98}.
In mono-modal logic, $I$ is a singleton, and the (only) index of the connectives is omitted for 
simplicity.
Given a signature $\Sigma$ of predicate and function symbols, the set of FOMML terms over 
$\Sigma$ is defined as for FOL. 
FOMML formulae over $\Sigma$ are given by the following abstract syntax:
\begin{minipage}[h]{\textwidth}
\begin{align*}
\varphi, \psi ::= p(t_1, \ldots, t_n) \mid t_1 = t_2 \mid &\neg \varphi \mid \varphi \land \psi \mid \varphi \lor \psi \mid \varphi \rightarrow \psi \mid \varphi \leftrightarrow \psi \mid \\ &\forall x.\, \varphi \mid \exists x.\, \varphi \mid \Box_i \varphi \mid \Diamond_i \varphi 
\end{align*}
\end{minipage}
where $t_i$ is a FOMML term, $p$ is a predicate symbol of arity $n$, $x$ is a variable, and
$i \in I$. 
As usual, existential quantification is dual to universal quantification, i.e.,
$\exists x.\, \varphi := \neg\forall x.\, \neg \varphi$, and $t_1 \neq t_2$ is syntactic sugar for 
$\neg(t_1 = t_2)$.
In normal modal logics the box and diamond connectives duals , i.e., 
$\Diamond_i\varphi := \neg\Box_i\neg\varphi$, $i \in I$.

The modal logic \textbf{K} is the smallest mono-modal logic that includes all closed FOL 
tautologies, all instances of the axiom scheme \textbf{K} ($\Box(\varphi \rightarrow \psi) 
\rightarrow (\Box\varphi \rightarrow \Box\psi)$), and is closed with respect to modus ponens 
(from $\varphi \rightarrow \psi$ and $\varphi$ infer $\psi$) and necessitation (from $\varphi$ 
infer $\Box \varphi$).
Further axiom schemes are assumed for stronger modal logics \cite{Gar18,FM23}.
In the multi-modal logic setting these axioms schemes can be imposed independently for every 
indexed connective.
The relationship between well-known axiom schemes is represented by the so-called 
modal cube~\cite{Gar18}.

Kripke-complete normal modal logics such as \textbf{K} can alternatively be characterized
by Kripke structures~\cite{Kri63}.
For a FOMML language over signature $\Sigma$, a first-order Kripke structure is a tuple
$M = (W, \mathcal{R}, \mathcal{D}, \mathcal{I})$, where $W$ is a non-empty set of worlds,
$\mathcal{R} = \{ R_i \}_{i \in I}$, $R_i \subseteq W \times W$ is the indexed accessibility 
relation between worlds, $\mathcal{D} = \{D_w\}_{w \in W}$ is the non-empty domains,
and $\mathcal{I} = \{ I_w \}_{w \in W}$ interprets the symbols of $\Sigma$ in world $w \in W$.
For stronger modal logics, restrictions may be imposed on each $R_i$, e.g., a \textbf{D} modality 
is characterized by the class of Kripke frames where $R_i$ is serial.
Similar correspondence results exist for the other common modal logics~\cite{Gar18}.
In Kripke semantics the truth of a formula $\phi$ with respect to $M$ and a world $w \in W$,
written $M,w \models \phi$, is defined as usual~\cite{FM23}.

Variants of FOMML semantics are discussed in the literature:
\begin{packed_itemize}
\item Quantification semantics can be specialized by domain restrictions \cite{FM23}:
      In \emph{constant domains} semantics (also called possibilist quantification), all domains 
      coincide, i.e., $D_w = D_v$ for all worlds $w,v \in W$.
      In \emph{cumulative domains} semantics, the domains are restricted such that 
      $D_w \subseteq D_v$ whenever $(w,v) \in R_i$, for all $w,v \in W$ and $i \in I$.
      In \emph{decreasing domains} semantics, the domains are restricted such that 
      $D_v \subseteq D_w$ whenever $(w,v) \in R_i$, for all $w,v \in W$ and $i \in I$.
      In \emph{varying domains} semantics (also called actualist quantification) no restriction 
      is imposed on the domains.
      These semantic restrictions can be equivalently characterized in a proof-theoretic way 
      using the (converse) Barcan formulae \cite{Bar46,FM23}.
\item Constancy restrictions of the interpretation of constant and function symbols across 
      different worlds can be applied.
      The interpretation $\mathcal{I}$ is \emph{rigid} if $I_w(c) = I_v(c)$ and $I_w(f) = I_v(f)$ 
      for each constant symbol $c \in \Sigma$ and function symbol $f \in \Sigma$, and all 
      $w,v \in W$.
      The interpretation is \emph{flexible} if this is not the case.
      This distinction is also referred to as rigid vs.\ flexible (or world-dependent) 
      \emph{designation}. 
      Details are discussed in \cite{FM23}.
\item The interpretation of terms can be constrained. 
      \emph{Local terms}~\cite{RO12} require that every ground term $t$ is interpreted in $w$ as 
      a domain element that exists in $w$, i.e., $I_w(c) \in D_w$ and 
      $I_w(f)(d_1, \ldots, d_n) \in D_w$, for every constant symbol $c \in \Sigma$, every 
      function symbol $f \in \Sigma$ of arity $n$, domain elements $d_1,\ldots,d_n \in D_w$,
      for every $w \in W$.
      In contrast, \emph{global terms} may be interpreted as any domain element 
      from $\bigcup_{w\in W} D_w$.
\end{packed_itemize}

Following Fitting and Mendelsohn~\cite{FM23}, consequence of a formula $\varphi$ in FOMML
is defined with respect to a set of global assumptions $G$ and a set of local assumptions $L$,
written $G \models L \rightarrow \varphi$. 
A Kripke structure $M$ globally satisfies a set of formulae $G$ iff $M,w \models \gamma$ for every 
$\gamma \in G$ and for every $w \in W$.
Then $G \models L \rightarrow \varphi$ iff for every Kripke structure $M$ that globally satisfies 
$G$ it holds that for every $w \in W$, if $M,w \models \lambda$ for every $\lambda \in L$ then 
$M,w \models \varphi$.
If $G$ is empty, this reduces to the common notion of local consequence.
(Local consequence and local assumptions should not be confused with local terms: the former speaks 
about the consequence relation, the latter about restrictions on the interpretation of terms.)

FOMML as given here deviates from some other descriptions of first-order modal logic in 
the literature, in so far as different descriptions usually restrict one or more of the semantic 
dimensions.
FOMML as given here is designed to be as general as possible, in order to incorporate specific 
modal logic variants as special cases.
For example, Fitting and Mendelsohn define their first-order modal logic without constant and 
functions symbols when focusing on their logic without predicate abstraction. 
This is easily simulated in FOMML by choosing $\Sigma$ accordingly.
Similarly, the modal logics used by the QMLTP assume rigid resignation and local consequence,
which can also be expressed in FOMML.

The typed first-order multi-modal logic used in the TPTP World is a typed variant of FOMML, with 
variables, function symbols, and predicate symbols having an assigned type signature. 
The type system is the same as for the classical typed first-order logic of the TPTP (see 
Section~\ref{ClassicalTPTPLanguages}).
Since the extension from untyped to typed FOMML is conceptually simple but technically cumbersome, 
it is not introduced here.

For higher-order modal logics (HOML), various logical systems have been proposed in the 
literature~\cite{Mon70,Mon73,Bre72,Gal75,Fit02}.  
Unfortunately, there is no immediate obvious generalization of these systems available, so that
no formal reference semantics of higher-order modal logics is introduced here. 
The TPTP syntax for HOML described in Section~\ref{NonClassicalTPTPLanguages} is general enough 
to represent many of the systems, but no reference semantics has been selected for the TPTP 
World. 
At this point it is up to the ATP system to choose what HOML system to implement. 

%--------------------------------------------------------------------------------------------------
\section{The Classical TPTP Languages}
\label{ClassicalTPTPLanguages}

The TPTP languages are human-readable, machine-parsible, flexible, and extensible languages,
suitable for writing both ATP problems and solutions.
The languages are:
\begin{packed_itemize}
\item First-order clause normal form (CNF)~\cite{SS98-JAR}, and full first-order form 
      (FOF)~\cite{Sut09}.
\item Typed first-order form (TFF), which adds types and type signatures. 
      TFF has monomorphic (TF0) and polymorphic (TF1) variants~\cite{SS+12,BP13-TFF1}.
\item Typed extended first-order form (TXF), which adds Boolean terms, Boolean variables as 
      formulae, tuples, conditional expressions, and let expressions~\cite{SK18}. 
      TXF has monomorphic (TX0) and polymorphic (TX1) variants.
\item Typed higher-order form (THF), which adds higher-order notions including curried type 
      declarations, lambda terms, partial application, and connectives as terms~\cite{SB10,KSR16}.
      THF has monomorphic (TH0) and polymorphic (TH1) variants.
\end{packed_itemize}
An overview that is relevant to this paper is given next.
The detailed syntax of the languages is given in an extended BNF\footnote{%
\label{BNF}\href{https://www.tptp.org/TPTP/SyntaxBNF.html}{\tt www.tptp.org/TPTP/SyntaxBNF.html}}~\cite{VS06}.

TPTP problems and solutions are built from {\em annotated formulae} of the form:
\begin{center}
{\em language}{\tt (}{\em name}{\tt ,}
{\em role}{\tt ,}
{\em formula}{\tt ,}
{\em source}{\tt ,}
{\em useful\_info}{\tt )}.
\end{center}
The {\em language}s supported are \smalltt{cnf}, \smalltt{fof}, \smalltt{tff}, and \smalltt{thf}.
The {\em role}, e.g., \smalltt{axiom}, \smalltt{lemma}, \smalltt{conjecture}, defines the 
use of the formula.
In a {\em formula}, terms and atoms follow Prolog conventions -- functions and predicates start 
with a lowercase letter or are {\tt '}single quoted{\tt '}, and variables start with an uppercase 
letter.
The languages also support interpreted symbols that either start with a {\tt \$}, e.g., the 
truth constants \smalltt{\$true} and \smalltt{\$false}, or are composed of 
non-alphabetic characters, e.g., integer/rational/real numbers such as 27, 43/92, -99.66.
The logical connectives in the TPTP language are
{\tt !>}, {\tt ?*}, {\tt @+}, {\tt @-}, {\tt !}, {\tt ?}, {\tt {\mytilde}}, {\tt |}, {\tt \&}, 
{\tt =>}, {\tt <=}, {\tt <=>}, and {\tt <{\mytilde}>}, for 
$\Pi$, $\Sigma$, choice (indefinite description), definite description,
$\forall$, $\exists$, $\neg$, $\vee$, $\wedge$, $\Rightarrow$, $\Leftarrow$, $\Leftrightarrow$, 
and $\oplus$ respectively.
Equality and inequality are expressed as the infix predicates {\tt =} and {\tt !=}.
The {\em source} and {\em useful\_info} are optional, to provide extra-logical information about 
the formula and its origin.

The typed first-order form (TFF) language supports types and type declarations.
Predicate and function symbols can be declared before their use, with type signatures that 
specify the types of their arguments and result.
Two TPTP defined types are available, {\tt \$i} for individuals, and {\tt \$o} for booleans.
User defined types can be introduced as being of the ``type'' {\tt \$tType}.
An expression {\tt ($t_1$,*\ldots*\,$t_n$)\,>\,\$o} is the type of an $n$-ary predicate, where 
the $i$-th argument is of type $t_i$.
Analogously, an expression {\tt ($t_1$\,*\ldots*\,$t_n$)\,>\,$t$} is the type of a function
that returns a term of type $t$.
TFF also supports arithmetic (which requires types, i.e., arithmetic is not supported in CNF or 
FOF), with defined types {\tt \$int}, {\tt \$rat}, {\tt \$real}, and a suite of interpreted 
arithmetic functions and predicates\footnote{%
\href{https://tptp.org/UserDocs/TPTPLanguage/TPTPLanguage.shtml\#ArithmeticSystem}{\tt tptp.org/UserDocs/TPTPLanguage/TPTPLanguage.shtml\#ArithmeticSystem}}.
A useful feature of TFF is default typing for symbols that are not explicitly declared:
predicates default to {\tt (\$i\,*\ldots*\,\$i)\,>\,\$o}, and
functions default to {\tt (\$i\,*\ldots*\,\$i)\,>\,\$i}.
This allows TFF to effectively degenerate to untyped FOF.

The typed extended first-order form (TXF) augments TFF with FOOL constructs:
formulae of type {\tt \$o} as terms; variables of type {\tt \$o} as formulae; tuples; 
conditional (if-then-else) expressions; and let (let-defn-in) expressions.
Figure~\ref{TXFExample} shows an example of a monomorphic typed extended first-order (TX0) 
annotated formula with its type declarations.
It expresses one of the axioms of the ``Knights and Knaves'' puzzles~\cite{Smu78}, that for every 
inhabitant {\tt I} and utterance {\tt S}, if {\tt I} is a knave (knaves always lie) and {\tt I}
says {\tt S} then {\tt S} is not true.
(This is an explicitly typed variant of the TPTP problem {\tt PUZ081\_8.p}\footnote{%
\href{https://tptp.org/cgi-bin/SeeTPTP?Category=Problems&Domain=SET&File=PUZ081_8.p}{\tt tptp.org/cgi-bin/SeeTPTP?Category=Problems\&Domain=SET\&File=PUZ081\_8.p}}.)
TXF provides the basis for the non-classical typed extended first-order form (NXF) introduced
in this paper.

\begin{figure}[h!]
\small
\setstretch{0.9}
\begin{verbatim}
%---------------------------------------------------------------------
tff(inhabitant_type,type, inhabitant: $tType).
tff(is_knave_decl,type,   is_knave: inhabitant > $o).
tff(says_decl,type,       says: (inhabitant * $o) > $o).

tff(knaves_lie,axiom,
    ! [I: inhabitant,S: $o] : 
      ( ( is_knave(I) & says(I,S) ) => ~ S ),
    file('PUZ081_8.p',knaves_lie),
    [description('Knaves always lie'), relevance(0.9)]).
%---------------------------------------------------------------------
\end{verbatim}
\caption{First-order types and an axiom in TXF.}
\label{TXFExample}
\end{figure}

\vspace*{1em}
The typed higher-order form (THF) is used for HOL. 
THF includes type declarations in curried form, $\lambda$-terms with a binder symbol {\tt \verb|^|}
for $\lambda$, explicit application with {\tt @}, and quantification over variables of any type.
THF does not admit default typing -- all symbol types must be declared before use.
Figure~\ref{THFExample} shows an example of a monomorphic typed higher-order (TH0) annotated 
formula.
It expresses the well-known argument of Cantor's proof that a powerset of a set
has a strictly larger cardinality than the set itself (here in the surjective variant).
This is the TPTP problem {\tt SET557\^{}1.p}\footnote{%
\href{https://tptp.org/cgi-bin/SeeTPTP?Category=Problems&Domain=SET&File=SET557^1.p}{\tt tptp.org/cgi-bin/SeeTPTP?Category=Problems\&Domain=SET\&File=SET557\^{}1.p}}.

\begin{figure}[h!]
\small
\setstretch{0.9}
\begin{verbatim}
%---------------------------------------------------------------------
thf(surjectiveCantorThm,conjecture,
    ~ ? [G: $i > $i > $o] :
      ! [F: $i > $o] :
      ? [X: $i] :
        ( ( G @ X ) = F ) ).
%---------------------------------------------------------------------
\end{verbatim}
\caption{Higher-order conjecture in THF.}
\label{THFExample}
\end{figure}

%--------------------------------------------------------------------------------------------------
\section{The Non-classical TPTP Languages}
\label{NonClassicalTPTPLanguages}

The non-classical typed form (NTF) family of TPTP languages were designed with the following 
principles in mind: (i)~syntactic consistency with the underlying classical languages, (ii)~a 
uniform syntax for a wide range of non-classical logics and connectives, and (iii)~requiring 
minimal changes to existing parsing and reasoning software.
The NTF languages add non-classical connectives, and a syntax for specifying the logic in which 
the problem is formulated.
NTF has two top level variants based on the classical typed extended first-order (TXF) and
typed higher-order (THF) languages.
They are the non-classical typed extended first-order form (NXF), and the non-classical 
typed higher-order form (NHF).
As with TXF and THF, NXF and NHF have monomorphic (NX0, NH0) and polymorphic (NX1, NH1) 
variants.
All constructs of the underlying TXF and THF languages are available in the NXF and NHF languages.

Figure~\ref{NX0Example} shows an alethic modal logic problem in NX0, and Figure~\ref{NH0Example} 
shows a higher-order conjecture in NHF that extends the conjecture from Figure~\ref{NX0Example}.
These figures provide running examples for the details that are explained in 
Sections~\ref{NonClassicalConnectives}, \ref{LogicSpecification}, and~\ref{QMLinNTF}.

\begin{figure}[h!]
\small
\setstretch{0.9}
\begin{verbatim}
%------------------------------------------------------------------------
tff(semantics,logic,
    $alethic_modal ==
      [ $domains == $constant,
        $designation == $rigid,
        $terms == $global,
        $modalities == $modal_system_M ] ).

tff(person_type,type,     person: $tType ).
tff(product_type,type,    product: $tType ).
tff(alex_decl,type,       alex: person ).
tff(leo_decl,type,        leo: product ).
tff(advisor_of_decl,type, advisor_of: person > person ).
tff(work_hard_decl,type,  work_hard: ( person * product ) > $o ).
tff(gets_rich_decl,type,  gets_rich: person > $o ).

tff(work_hard_to_get_rich,axiom,
    ! [P: person] :
      ( ? [R: product] : work_hard(P,R)
     => ( {$possible} @ ( gets_rich(P) ) ) ) ).

tff(not_all_get_rich,axiom,
    ~ ? [P: person] : ( {$necessary} @ ( gets_rich(P) ) ) ).

tff(one_rich,axiom,
    ! [P: person,O: person] :
      ( ( gets_rich(P) & O != P )
     => ( {$necessary} @ ( ~ gets_rich(O) ) ) ) ).

tff(no_self_advising,axiom,
    ! [P: person] : P != advisor_of(P) ).

tff(alex_works_on_leo_here,hypothesis,
    work_hard(alex,leo) ).

tff(alex_advisor_works_on_leo_here,hypothesis,
    work_hard(advisor_of(alex),leo) ).

tff(someone_gets_rich_but_not_advisor,conjecture,
    ? [P: person] :
      ( {$possible} @ ( gets_rich(P) & ~ gets_rich(advisor_of(P)) ) ) ).
%------------------------------------------------------------------------
\end{verbatim}
\caption{A first-order modal logic problem in NX0.}
\label{NX0Example}
\end{figure}

\begin{figure}[h!]
\small
\setstretch{0.9}
\begin{verbatim}
%------------------------------------------------------------------------
thf(someone_gets_rich_but_not_advisor,conjecture,
    ? [P: person] :
    ? [F: person > person] :
      ( {$possible} @ ((gets_rich @ P)) & ~ (gets_rich @ (F @ P)) ) ).
%------------------------------------------------------------------------
\end{verbatim}
\caption{A higher-order modal logic conjecture in NH0.}
\label{NH0Example}
\end{figure}

%--------------------------------------------------------------------------------------------------
\subsection{Non-classical Connectives}
\label{NonClassicalConnectives}

The NTF non-classical connectives are enclosed in braces, in the form 
{\tt \{}{\em connective\_name}{\tt \}}.
The {\em connective\_name} is either a TPTP defined symbol starting with {\tt \$}, e.g.,
{\tt \{\$box\}}, or a system defined symbol starting with {\tt \$\$}.
The meanings of TPTP defined symbols are documented in the TPTP, while the meaning of system 
defined symbols are known by the systems that support them.
System defined symbols allow the TPTP syntax to be used when experimenting with logics that have 
not been defined in the TPTP. 
Examples of simple non-classical connectives that can be represented in NTF include simple 
unary modal connectives $\Box$ and $\Diamond$ from modal logics, binary conditional connectives 
$>$ of conditional logics~\cite{ER21}, and the binary linear logic connectives $\&$, 
\rotatebox[origin=c]{180}{$\&$} and $\multimap$~\cite{Gir98}.

The non-classical connectives are applied explicitly using {\tt @}, in a mixed 
higher-order-applied/first-order-functional style\footnote{%
This slightly unusual form was chosen to reflect the first-order functional style, and 
the explicit application using {\tt @} can be parsed in Prolog -- a long standing 
principle of the TPTP languages~\cite{SZS04}.}:
\begin{itemize}
\item In NXF, {\tt \verb|{|}{\em connective\_name}{\tt \verb|}|} {\tt @} {\tt (}{\em arg$_1$}{\tt ,}{\em ...}{\tt ,}{\em arg$_n$}{\tt )}
      is a formula, where each {\em arg$_i$} is an NXF formula (see Figure~\ref{NX0Example} for 
      examples of the syntax).
\item In NHF, {\tt \verb|{|}{\em connective\_name}{\tt \verb|}|} {\tt @} {\em arg$_1$} {\tt @} {\em ...} {\tt @} {\em arg$_n$}
      is a formula, where each {\em arg$_i$} is an NHF formula (see Figure~\ref{NH0Example} for
      an example of the syntax).
\end{itemize}

Despite their functional appearance, NTF connectives are different from classical connectives.
Firstly, they can have any arity, and secondly they can be parameterized with formulae and terms 
to implement more complex non-classical connectives.
Figure~\ref{NX0DefaultsExample} shows an illustrative example in an epistemic logic extended with 
a non-monotonic conditional {\tt \{\$usually\}}.
The axiom {\tt alex\_knows\_birds\_fly} says that if it is common knowledge among agents 
{\tt geoff} and {\tt chris} that every bird usually flys, then {\tt alex} believes that all birds 
fly.
Note how {\tt \{\$usually\}} is a binary connective with the arguments {\tt bird(X)} and 
{\tt fly(X)}, while {\tt \{\$believes\}} is a unary connective with the argument 
{\tt !\,[Y]\,:\,(\,bird(Y)\,=>\,fly(Y)\,)}.
The general form with parameters is
{\tt \{}{\em connective\_name}{\tt (}{\em param$_1$}{\tt ,}{\em \ldots}{\tt ,}{\em param$_n$}{\tt )}{\tt \}}.
If the connective is indexed the index is given as the first argument as a {\tt \#}-prefixed
constant (uninterpreted constant, number, or TPTP defined constant), e.g., 
{\tt \{\$dia(\#francis\_of\_assisi)\}\,@\,(\,do\_the\_impossible\,)}.
All other parameters are key-value pairs of the form 
{\em parameter\_name}~{\tt :=}~{\em parameter\_value}, where the {\em parameter\_name} is a 
constant, and the {\em parameter\_value} is a term or formula.
In Figure~\ref{NX0DefaultsExample} the unary {\tt \{\$common\}} connective is parameterized by the 
key-value pair {\tt agents:=[geoff,chris]}.
The full specification of the NTF connective syntax and their use in formulae is in the BNF.

\begin{figure}[h!]
\small
\setstretch{0.9}
\begin{verbatim}
%---------------------------------------------------------------------
tff(alex_knows_birds_fly,axiom,
    ( {$common(agents:=[geoff,chris])} @ 
      ! [X: $i] : ( {$usually} @ ( bird(X) , fly(X) ) )
   => ( {$believes(#alex)} @ ( ! [Y] ( bird(Y) => fly(Y) ) ) ) ) ).

tff(tweety_is_bird,axiom, bird(tweety) ).

tff(tweety,conjecture, fly(tweety) ).
%---------------------------------------------------------------------
\end{verbatim}
\caption{A first-order modal logic problem with parameterized connectives and default typing, in TXF.}
\label{NX0DefaultsExample}
\end{figure}

Examples of parameterized non-classical connectives that can be represented in NTF include
indexed unary modal connectives such as $\Box_i$ from multi-modal logics, unary agent-relative 
knowledge and agent-group common knowledge connectives such as $K_a$ and $C_{\{a,b,c\}}$ from 
epistemic logics~\cite{vDH15}, unary term-(sequence) indexed connectives such as 
$[t_1, \ldots, t_n]$ from term-(sequence) model logics~\cite{FTV01,SSY19}, unary connectives such 
as $[P]$ from dynamic logics~\cite{HKT01} that are indexed by Kleene albgera expressions that 
represent complex programs, and mixtures of these as in dynamic epistemic logics~\cite{vvK07}.

The TPTP language extension for non-classical connectives is arguably conservative. 
A simpler and more concise syntax could be envisioned for individual logics, e.g. normal modal 
logic.
However, NTF aims at providing a future-proof syntax that can represent logics with arbitrary 
$n$-ary connectives that can carry non-trivial parameters.
The NTF syntax also makes a clear distinction between the object logic and meta-logical components, 
e.g., the expression {\tt \{\$believes(\#alex)\}\,@\,!\,[Y]\,(\,bird(Y)\,=>\,fly(Y)\,)} is clearly 
presented as the unary connective {\tt \{\$believes(\#alex)\}} with a meta-logical parameter 
{\tt \#alex}, which is applied to the argument {\tt !\,[Y]\,(\,bird(Y)\,=>\,fly(Y)\,)}.
This contrasts with the syntactically simpler option of making {\tt \$believes} a binary predicate 
that is applied to the index parameter and the argument.
The NTF syntax thus addresses at least two problems of syntactically simpler approaches:
(i)~In the NTF syntax, connectives can have parameters, while the simpler approach might need 
connectives with parameters as arguments.
(ii)~In the NTF syntax, parameters such as {\tt \#alex} are not necessarily part of the object
logic (neither terms nor formulae) and in that case do not be need to be given types. 
In a simpler approach it might be necessary to declare the types of parameters in order for their 
occurrences at the object level to be well-typed.
The NTF syntax is a trade-off between conciseness and generality, and allows for a uniform 
representation of non-classical connectives with arbitrary arity.

As was noted in Section~\ref{ClassicalTPTPLanguages}, the default typing rules of TFF and TXF, 
and hence also of NXF, allows the languages to degenerate to untyped languages. 
This is useful as many non-classical logics are untyped.
In Figure~\ref{NX0DefaultsExample}, {\tt bird} and {\tt fly} default to predicates of type
{\tt \$i > \$o}, {\tt tweety} defaults to a constant of type {\tt \$i}, and {\tt Y}
defaults to a variable of type {\tt \$i}. 

%--------------------------------------------------------------------------------------------------
\subsection{Logic Specifications}
\label{LogicSpecification}

In the world of non-classical logics the intended logic cannot always be inferred from the
language used for the formulae -- the same language can be used for different logics, e.g., 
modal logics and intuitionistic logics share the same language but admit different inferences.
It is therefore necessary to provide (meta-)information that specifies the logic to be used.
In NTF a TPTP annotated formula with the role {\tt logic} is used for this, with a 
\textit{logic specification} as the formula. 
A logic specification consists of a logic name with a list of properties, in the form: \\
\hspace*{1cm}{\tt tff(}{\em name}{\tt,logic,}{\em logic\_name} {\tt ==} {\em properties}{\tt ).} \\
where {\em properties} is a {\tt \verb|[]|} bracketed list of key-value identities in the form: \\
\hspace*{1cm}{\em property\_name} {\tt ==} {\em property\_value} \\
Each {\em property\_name} is a TPTP defined symbol or a system defined symbol,
and each {\em property\_value} is either a term of the language (often a defined constant) or a 
{\tt \verb|[]|} bracketed list that might start with a term and otherwise contains key-value 
identities.
If the first element of a {\em property\_value} list is a term then that is the default value 
for all cases that are not specified by the following key-value identities.
The order in which the properties are specified does not matter.
Figure~\ref{NX0Example} shows a simple example for alethic modal logic, and 
Figure~\ref{complexLogic} shows a more complex example.

An NTF file must have one logic specification, and it normally comes first in the file. 
A file without a logic specification defaults to classical logic, and it is therefore an error 
to use a non-classical connective without a preceding logic specification.
In order to reduce misunderstandings, there are no general default values for properties --
it is an error if a logic specification does not specify all relevant properties.

The full grammar for logic specifications is in the BNF.
The grammar is quite unrestrictive, and allows for complex specifications, e.g., 
arbitrary formulae can be used as {\em property\_value}s.
It is flexible enough to be used for many different logics, users can create specifications 
for logics that are not defined in the TPTP, and it is possible to specify the same logic in 
different ways.

%--------------------------------------------------------------------------------------------------
\section{Quantified Multi-Modal Logics in NTF}
\label{QMLinNTF}

The first major use case of NTF in the TPTP World is quantified normal multi-modal logic, as 
introduced in Section~\ref{Preliminaries}.\footnote{%
The second logic family currently supported by the TPTP is Prior's instant-based temporal logics 
TL, named {\tt \$temporal\_instant} in the TPTP.
Technically, the TL logic family is a specific modal logic family with two different modalities
-- suggestively named $P$ and $F$ for past and future, respectively -- that are constrained by 
additional interaction axiom schemes~\cite{GR22}.
TL is described in more detail at 
\href{https://tptp.org/UserDocs/TPTPLanguage/TPTPLanguage.shtml\#NonClassicalLogics}{\tt tptp.org/UserDocs/TPTPLanguage/TPTPLanguage.shtml\#NonClassicalLogics}}
The multi-modal connectives $\Box_i$ and $\Diamond_i$ are provided as {\tt \{\$box(\#i)\}} and 
{\tt \{\$dia(\#i)\}}.
If only one modality is required, the unindexed variants {\tt \{\$box\}} and {\tt \{\$dia\}} are 
normally used.
There are also short form unary connectives for unparameterised {\tt \{\$box\}} and
{\tt \{\$dia\}}: {\tt [.]} and {\tt <.>}, e.g., {\tt \{\$box\}\,@\,(p)} can be written 
{\tt [.]\,p}.

In addition to the purely syntactic flavor of modal logic where the modal connectives 
{\tt \{\$box\}} and {\tt \{\$dia\}} do not have any intended interpretation, the TPTP provides 
variants where the modal connectives are named according to their intended interpretation.
This gives five modal logic families:
\begin{packed_itemize}
\item {\tt \$modal} -- Modal logic, with the connectives {\tt \{\$box\}} and {\tt \{\$dia\}}.
\item {\tt \$alethic\_modal} -- Modal logic with an alethic interpretation, with the
      connectives {\tt \{\$necessary\}} and {\tt \{\$possible\}}.
\item {\tt \$deontic\_modal} -- Modal logic with an deontic interpretation, with the
      connectives {\tt \{\$obligatory\}} and {\tt \{\$permissible\}}.
\item {\tt \$epistemic\_modal} -- Modal logic with an epistemic interpretation, with the
      connectives {\tt \{\$knows\}} and {\tt \{\$canKnow\}}.
\item {\tt \$doxastic\_modal} -- Modal logic with a doxastic interpretation, with the
      connectives {\tt \{\$believes\}} and {\tt \{\$canBelieve\}}.
%\item {\tt \$temporal\_instant} -- Instant-based temporal logics, a system in which time is 
%      considered as a discrete flow of instants, with temporal connectives
%      {\tt \{\$P\}}, {\tt \{\$F\}}, {\tt \{\$G\}}, and {\tt \{\$H\}}.
\end{packed_itemize}
The logics {\tt \$alethic\_modal}, {\tt \$deontic\_modal}, {\tt \$epistemic\_modal} and 
{\tt \$doxastic\_modal} are simply syntactic variants of the generic {\tt \$modal} logic, 
with connectives renamed for readability. 
The remainder of this section applies equally to all the variants.

To differentiate between local and global assumptions in a problem, the roles of the TPTP
annotated formulae are used.
The role {\tt hypothesis} indicates that the formula is a local assumption, and
the role {\tt axiom} indicates that the formula is a global assumption. 
A formula with role {\tt conjecture} is the one to be proved, i.e., the right-hand side
of the consequence relation.
{\em Subroles} can be used to override the local/global defaults.
A formula with subrole {\tt -local} is always a local assumption, e.g., {\tt axiom-local} is 
local instead of global.
Similarly, a formula with subrole {\tt -global} is always a global assumption.
% , and a formula with the role {\tt conjecture-global} is to be proved globally.
The assumptions {\tt alex\_works\_on\_leo\_here} and {\tt alex\_advisor\_works\_on\_leo\_here} 
in Figure~\ref{NX0Example} are local while the other axioms are global.

%--------------------------------------------------------------------------------------------------
\subsection{Logic Specifications in {\tt \$modal}}
\label{LogicSpecModal}

A {\tt \$modal} logic specification documents four properties that characterize the modal logic 
to be used. 
It has the following form:
\begin{lstlisting}[basicstyle=\ttfamily,columns=fullflexible,keepspaces=true,escapechar={\#}]
tff(semantics,logic,
    $modal == [ $domains     == #\normalfont\textit{domains\_specification}#,
                $designation == #\normalfont\textit{designation\_specification}#,
                $terms       == #\normalfont\textit{terms\_specification}#,
                $modalities  == #\normalfont\textit{modalities\_specification}# ] ).
\end{lstlisting}
The first three properties describe semantic constraints applied to the Kripke model structures,
following the definitions in Section~\ref{Preliminaries}.
The last property is a proof theoretic property.
The properties have the following meanings:
\begin{packed_itemize}
\item {\tt \$domains} specifies the constancy restrictions on the individual domains $D_w$, 
      $w\in W$, across different accessible worlds.
\item {\tt \$designation} specifies whether symbols are interpreted rigidly or flexibly by $I_w$ 
      across different worlds $w \in W$.
\item {\tt \$terms} specifies whether terms must be interpreted as domain elements in the local 
      world $w \in W$, or can be interpreted as any domain element from any world $w \in W$.
\item {\tt \$modalities} specifies the character of each connective as a frame property or modal 
      axiom schemes.
\end{packed_itemize}
Table~\ref{modal_properties} shows the possible values for \textit{domain\_specification}, 
\textit{designation\_specification} and \textit{terms\_specification}, with meanings per their 
naming.
For the {\tt \$modalities} property, the \textit{modalities\_specification} is either a 
\textit{mono\_modal\_spec} or a \textit{multi\_modal\_spec}.
A \textit{mono\_modal\_spec} specifies a mono-modal logic, i.e., a modal logic with only 
one {\tt \{\$box\}} and {\tt \{\$dia\}} connective.
It is either a parameter value {\tt \$modal\_system\_$X$}, where {\tt X} denotes the modal 
system name to be used, or a list {\tt [\$modal\_axiom\_$X$, \$modal\_axiom\_$Y$, \ldots]}
naming the modal axiom schemes to use.
Table~\ref{ModalSystems} lists the possible values for {\tt \$modal\_system\_$X$}, with their 
relationships to the individual axiom schemes. 
Table~\ref{ModalAxioms} lists the possible values for {\tt \$modal\_axiom\_$X$}, with their 
meanings.
A \textit{multi\_modal\_spec} specifies a multi-modal logic, i.e., a modal logic with multiple
{\tt \{\$box(\#i)\}} and {\tt \{\$dia(\#i)\}} connectives.
It is a comma-separated list of 
\textit{modal\_connective}~{\tt ==}~\textit{mono\_modal\_spec} pairs of the form:\\
\hspace*{1em}{\tt [ {\$box(\#$i_1$)} == \textit{mono\_modal\_spec},
{\$box(\#$i_2$)} == \textit{mono\_modal\_spec}, ...]}\\
specifying the desired properties of the connectives for each index. 
Optionally, if the first entry of the list is a \textit{mono\_modal\_spec}, i.e., 
without a modal connective key, that \textit{mono\_modal\_spec} is used as the default value
for all connectives that are not specified by an explicit key-value entry.
As {\tt \{\$box\}} and {\tt \{\$dia\}} are dual, specifying the modality of one implicitly 
specifies the same modality for the other.

\begin{table}[h]
\caption{Logic specification properties of {\tt \$modal}.}
\label{modal_properties}
\centering
\begin{tabular}{p{.10\textwidth}|p{.50\textwidth}}
{\bf Property} & {\bf Possible values} \\[.2em]
\hline
{\tt \$domains} & {\tt \$constant}, {\tt \$varying}, {\tt \$cumulative}, {\tt \$decreasing} \\[0.2em]
\rowcolor{mygray}
{\tt \$designation}      & {\tt \$rigid}, {\tt \$flexible} \\[0.2em]
{\tt \$terms} & {\tt \$global}, {\tt \$local} \\[0.2em]
%\rowcolor{mygray}
%{\tt \$modalities}     & {\tt \$modal\_system\_X} \\
%\rowcolor{mygray}      & {\tt X} $\in$ \{{\tt K}, {\tt KB}, {\tt K4}, {\tt K5}, {\tt K45}, 
%                         {\tt KB5}, {\tt D}, {\tt DB}, {\tt D4}, {\tt D5}, {\tt D45}, {\tt T},
%                         {\tt B}, {\tt S4}, {\tt S5}, {\tt S5U}\} \\
%\rowcolor{mygray}      & {\em or a list of axioms} \\
%\rowcolor{mygray}      & {\tt [\$modal\_axiom\_X$_1$, \$modal\_axiom\_X$_2$, ...]} \\
%\rowcolor{mygray}      & {\tt X$_i$} $\in$ \{{\tt K}, {\tt T}, {\tt B}, {\tt D}, {\tt 4}, {\tt 5}, 
%                         {\tt CD}, {\tt BoxM}, {\tt C4}, {\tt C}\} \\
\hline
\end{tabular}
\end{table}

\begin{table}[htb]
\caption{The {\tt \$modal\_system\_$X$} values and their meanings.}
\label{ModalSystems}
\centering
\begin{tabular}{p{.10\textwidth}|p{.20\textwidth}|p{.30\textwidth}}
\bf System & {\bf TPTP name} & {\bf Axiom schemes} \\[.2em]
\hline
\bf K & {\tt \$modal\_system\_K} & $\{K\}$\\[0.2em]
\rowcolor{mygray}
\bf KB & {\tt \$modal\_system\_KB} & $\{K,B\}$ \\[0.2em]
\bf K4 & {\tt \$modal\_system\_K4} & $\{K,4\}$ \\[0.2em]
\rowcolor{mygray}
\bf K5 & {\tt \$modal\_system\_K5} & $\{K,5\}$ \\[0.2em]
\bf K45 & {\tt \$modal\_system\_K45} & $\{K,4,5\}$ \\[0.2em]
\rowcolor{mygray}
\bf KB5 & {\tt \$modal\_system\_KB5} & $\{K,B,5\}$  \\[0.2em]
\bf D & {\tt \$modal\_system\_D} & $\{K,D\}$ \\[0.2em]
\rowcolor{mygray}
\bf DB & {\tt \$modal\_system\_DB} & $\{K,D,B\}$ \\[0.2em]
\bf D4 & {\tt \$modal\_system\_D4} & $\{K,D,4\}$ \\[0.2em]
\rowcolor{mygray}
\bf D5 & {\tt \$modal\_system\_D5} & $\{K,D,5\}$ \\[0.2em]
\bf D45 & {\tt \$modal\_system\_D45} & $\{K,D,4,5\}$ \\[0.2em]
\rowcolor{mygray}
{\bf M} {\it or} {\bf T}  & {\tt \$modal\_system\_M} & $\{K,M\}$ \\[0.2em]
\bf B & {\tt \$modal\_system\_B} & $\{K,B\}$ \\[0.2em]
\rowcolor{mygray}
\bf S4 & {\tt \$modal\_system\_S4} & $\{K,M,4\}$ \\[0.2em]
\bf S5 & {\tt \$modal\_system\_S5} & $\{K,M,B,5\}$ \\[0.2em]
%\rowcolor{mygray}
%\bf S5U & {\tt \$modal\_system\_S5U} & $\{ \}$ \\[0.2em]
\hline
\end{tabular}
\end{table}

\begin{table}[htb]
\caption{The {\tt \$modal\_axiom\_$X$} values and their meanings.}
\label{ModalAxioms}
\centering
\begin{tabular}{p{.10\textwidth}|p{.20\textwidth}|p{.30\textwidth}}
\bf Name & {\bf TPTP name} & {\bf Axiom scheme} \\[.2em]
\hline
K & {\tt \$modal\_axiom\_K} & $\Box(\phi \to \psi) \to \Box \phi \to \Box \psi$ \\[0.2em]
\rowcolor{mygray}
M {\it or} T & {\tt \$modal\_axiom\_M} & $\Box\phi \to \phi$ \\[0.2em]
B & {\tt \$modal\_axiom\_B} &  $\phi \to \Box\Diamond\phi$\\[0.2em]
\rowcolor{mygray}
D & {\tt \$modal\_axiom\_D} & $\Box\phi \to \Diamond\phi$\\[0.2em]
4 & {\tt \$modal\_axiom\_4} & $\Box\phi\to\Box\Box\phi$\\[0.2em]
\rowcolor{mygray}
5 & {\tt \$modal\_axiom\_5} & $\Diamond\phi\to\Box\Diamond\phi$\\[0.2em]
CD & {\tt \$modal\_axiom\_CD} & $\Diamond\phi\to\Box\phi$\\[0.2em]
\rowcolor{mygray}
$\Box$M & {\tt \$modal\_axiom\_BoxM} & $\Box(\Box\phi\to\phi)$\\[0.2em]
C4 & {\tt \$modal\_axiom\_C4} & $\Box\Box\phi\to\Box\phi$\\[0.2em]
\rowcolor{mygray}
C & {\tt \$modal\_axiom\_C} & $\Diamond\Box\phi\to\Box\Diamond\phi$\\[0.2em]
\hline
\end{tabular}
\end{table}

The NX0 example problem in Figure~\ref{NX0Example} includes a logic specification that 
specifies that alethic modal logic is being used, the domains are constant -- the same in all 
worlds, the interpretation of symbols is rigid -- the same in all worlds, terms' denotation is 
global -- from any world, and all connectives have \textbf{M} modality.
Given this logic specification the conjectures from Figures~\ref{NX0Example} and~\ref{NH0Example} 
can both be proved.
However, if either the {\tt \$domains} is changed to {\tt \$varying}, or the 
{\tt \$modalities} is weakened to {\tt \$modal\_system\_D} or {\tt \$modal\_system\_K}, then 
the conjectures cannot be proved, for the following reasons:
(i)~In the constant domain setting all domains coincide and it is assured that the interpretation 
of {\tt alex} exists locally, serving as witness for {\tt P}.
If the {\tt \$domains} is changed to {\tt \$varying} then it is not known whether {\tt alex} 
exists in the local world (where the conjecture is being proved) because terms are global and 
{\tt alex} might be interpreted as a domain element that comes from only a different world.
As a result there might not exist a hard working domain element {\tt P} in the local world.
(ii)~The axiom scheme M makes all accessibility relations reflexive. 
If accessibility is not reflexive, it is not known whether the world in which the person {\tt P} 
gets rich also has an advisor who does not get rich. 
The necessary statement of formula {\tt one\_rich} might not apply in the world where {\tt alex} 
gets rich, because the latter world is not necessarily accessible from itself.

A more complex logic specification example is displayed in Figure~\ref{complexLogic}.
Here, the {\tt \$modal} logic has cumulative domains, flexible constants, and local terms.
It is a multi-modal logic with at least two indices for the connectives, {\tt \#1} and {\tt \#2}.
{\tt \{\$box(\#1)\}} and {\tt \{\$dia(\#1)\}} are \textbf{S5} connectives, and 
{\tt \{\$box(\#2)\}} and {\tt \{\$dia(\#2)\}} satisfy axiom schemes K, D and C4. 
All other modal connectives with other indices are \textbf{K} connectives, per the default 
value {\tt \$modal\_system\_K}.

\begin{figure}[hbt]
\small
\setstretch{0.9}
\begin{verbatim}
%------------------------------------------------------------------------
tff(semantics,logic,
    $modal ==
      [ $domains == $cumulative,
        $designation == $flexible,
        $terms == $local,
        $modalities == 
          [ $modal_system_K,
            {$box(#1)} == $modal_system_S5,
            {$box(#2)} == [$modal_axiom_K, $modal_axiom_D, 
                           $modal_axiom_C4] ] ] ).
%------------------------------------------------------------------------
\end{verbatim}
\caption{A multi-modal logic specification with a default {\tt \$modalities} value.}
\label{complexLogic}
\end{figure}

%--------------------------------------------------------------------------------------------------
\section{TPTP Infrastructure for NTF}
\label{TPTP}

%--------------------------------------------------------------------------------------------------
\subsection{The TPTP Problem Library}
\label{TPTPProblemLibrary}

The TPTP problem library~\cite{Sut17} is the de facto standard set of test problems for ATP 
systems.
The TPTP supplies the ATP community with a comprehensive library of test problems that
provides an overview and a simple, unambiguous reference mechanism. 
Since its first release in 1993 many researchers have used the TPTP problem library as an 
appropriate and convenient basis for ATP system research and development.
The TPTP problem library includes non-classical logic problems, written in the NTF language.

Problem files in the library have three parts: a header, optional includes, and annotated formulae. 
The header section provides information about the problem for users, formatted as comments.
The include section contains {\tt include} directives for axiom files, to avoid duplication of 
formulae in commonly used axiomatizations. 
Annotated formulae are described in Section~\ref{ClassicalTPTPLanguages}.

The header section of a problem file has four parts.
The first part identifies and describes the problem.
The second part provides information about occurrences of the problem in the literature and 
elsewhere. 
The third part provides semantic and syntactic characteristics of the problem. 
The last part contains comments and bugfix information.
The header section's fields are self-explanatory\footnote{%
e.g., \href{https://tptp.org/cgi-bin/SeeTPTP?Category=Problems&Domain=SYO&File=SYO886_1.111.p}{\tt tptp.org/cgi-bin/SeeTPTP?Category=Problems\&Domain=SYO\&File=SYO886\_1.111.p}}. 
The following are of particular interest here:
\begin{packed_itemize}
\item The {\tt Status} field gives the semantic status of the problem as an SZS ontology 
      value~\cite{Sut08-KEAPPA}.
      The most common values are {\tt Theorem} - the conjecture can be proved from the axioms,
      {\tt CounterSatisfiable} - the conjecture cannot be proved from the axioms,
      {\tt Unsatisfiable} - the formulae are unsatisfiable, and
      {\tt Satisfiable} - the formulae are satisfiable.
\item The {\tt Syntax} field provides statistics on the use of TPTP language in the problem.
\item The {\tt SPC} field gives the Specialist Problem Class (SPC) of the problem.
      Each SPC identifies a class of problems with the same recognizable logical, language, and 
      syntactic characteristics, which make the problems in the SPC homogeneous wrt ATP systems.
\item The {\tt Rating} field provides a numerical difficulty rating for the problem~\cite{SS01}.
      Ratings range from 0.00 (easy), through 0.01 to 0.99 (difficult), up to 1.00 (unsolved
      by any ATP system).
      % - problems that are solved by all non-subsumed ATP systems 
      % (not solving a subset of the problems solved by another system) known in the TPTP World),
      % from 0.01 and 0.99 (difficult) - problems that are solved by some non-subsumed ATP systems, 
      % and 1.00 (unsolved) - problems that are solved by no ATP systems.
      The ratings are computed wrt each SPC.
\end{packed_itemize}
Figure~\ref{NX0Header} shows these fields for the problem in Figure~\ref{NX0Example}. 
The conjecture of the problem is a {\tt Theorem} of the axioms. 
It is not a very difficult problem (rating 0.13).
The problem has four non-indexed and no indexed non-classical connectives
({\tt 4~\{.\};~~~0~\{\#\}}).
The problem is written in the monomorphic non-classical typed extended 
first-order form ({\tt NX0}), is a theorem ({\tt THM}), includes some equality ({\tt EQU}), 
and does not use arithmetic ({\tt NAR}).

\begin{figure}[h!]
\small
\setstretch{0.9}
\begin{verbatim}
%------------------------------------------------------------------------------
% Status   : Theorem
% Rating   : 0.13 v9.1.0
% Syntax   : Number of formulae    :   14 (   3 unt;   7 typ;   0 def)
%            Number of atoms       :   15 (   2 equ)
%            Maximal formula atoms :    4 (   2 avg)
%            Number of connectives :   17 (   5   ~;   0   |;   2   &)
%                                         (   0 <=>;   2  =>;   0  <=;   0 <~>)
%                                         (   4 {.};   0 {#})
%            Maximal formula depth :    6 (   3 avg)
%            Maximal term depth    :    2 (   1 avg)
%            Number of types       :    3 (   2 usr)
%            Number of type conns  :    4 (   3   >;   1   *;   0   +;   0  <<)
%            Number of predicates  :    5 (   2 usr;   2 prp; 0-2 aty)
%            Number of functors    :    3 (   3 usr;   2 con; 0-1 aty)
%            Number of variables   :    7 (   4   !;   3   ?;   7   :)
% SPC      : NX0_THM_EQU_NAR
%------------------------------------------------------------------------------
\end{verbatim}
\caption{The header for the NX0 problem in Figure~\ref{NX0Example}.}
\label{NX0Header}
\end{figure}

The NTF problems in the TPTP problem library v9.1.0 are all normal mono-modal logic problems, 
including problems from (the citations are just some examples) 
books~\cite{For94,FM23,Gir00,Sid10}, conference and journal papers~\cite{Rei92,FH+98,Sto00,PN+21}, 
and use cases~\cite{BW14-ECAI,MR22}.
There are:
\begin{packed_itemize}
\item 215 NTF problems (excluding the {\tt SYN000} problems that are designed only for parser 
      testing).
\item 194 NX0 problems and 21 NH0 problems (0 NX1 or NH1 polymorphic problems).
\item 52 propositional problems, 129 NX0 problems without equality, 13 NX0 problems 
      with equality, 9 NH0 problems without equality, and 12 NH0 problems with equality.
\item 154 theorems, 45 countersatisfiable problems, and 16 problems with unknown status.
\item 12 NX0 {\tt GRA} (graph theory) problems, 
      12 NX0 {\tt LCL} (logic calculi) problems, 
      15 NH0 {\tt PHI} (philosophy) problems,
      10 NX0 {\tt PLA} (planning) problems, 
      6 NH0 {\tt SYO} (syntactic) problems, 
      and 
      158 NX0 {\tt SYO} (syntactic) problems.
\item 86 problems with {\bf K} modality,
       4 with {\bf K4} modality,
       4 with {\bf K45} modality,
       8 with {\bf KB} modality,
      17 with {\bf D} modality,
       6 with {\bf D4} modality,
      21 with {\bf M} modality,
       1 with {\bf B} modality,
      38 with {\bf S4} modality, and
      30 with {\bf S5} modality.
\item For the 163 non-propositional problems, there are:
      \begin{packed_itemize}
      \item 68 problems with constant domains, 21 with cumulative domains, 22 with decreasing
            domains, and 52 with varying domains.
      \item 123 problems with rigid designation, 40 with flexible designation.
      \item 63 problems with local terms, 100 with global terms.
      \end{packed_itemize}
% \item 147 problems that were in TPTP release v9.0.0, and thus have ratings.
%       59 have rating 0.0, 33 have rating 0.01 to 0.99, and 55 have rating 1.00.
%       The average rating of the 147 problems is 0.48.
\end{packed_itemize}
\noindent
The TPTP problem library documentation provides summary information:
the {\tt TFFSynopsis}, {\tt THFSynopsis}, and {\tt OverallSynopsis} files give the numbers of 
non-classical logic problems, and the {\tt ProblemAndSolutionStatistics} file gives the numbers 
of non-classical connectives in each problem.

%--------------------------------------------------------------------------------------------------
\subsection{ATP Systems}
\label{ATPSystems}

ATP for non-classical logics is a well established endeavour, especially for propositional
non-classical logics, but there are significantly fewer ATP systems available than for classical 
logics.
The ATP systems for modal logics that we know of are KSP~\cite{NHD20,PN+21}, 
nanoCoP-M~\cite{Ott21}, MleanCoP~\cite{Ott14}, MetTeL2~\cite{TSK12}, Leo-III~\cite{SB21}, 
LoTREC~\cite{FF+01}, and MSPASS~\cite{HS00-TABLEAUX}.
Of those, KSP, nanoCoP-M, MleanCoP, and Leo-III have been integrated into the TPTP World.
KSP is limited to propositional problems, but covers all the modalities listed in 
Table~\ref{ModalAxioms}.
nanoCoP-M and MleanCoP are limited to first-order problems with only local terms, local axioms, 
non-decreasing domains, and {\bf D}, {\bf M}, {\bf S4}, and {\bf S5} modalities (notably excluding 
{\bf K}).
Leo-III offers almost full generality for NTF problems, with the only limitation being to rigid
designation.
A comparative study of the performance of these systems is in~\cite{SS+23}.

In order to support existing ATP systems that do not read TPTP NX0 or NH0 syntax, some translators 
have been implemented to convert NTF formulae to other systems' native syntaxes.
Thus far translators have been implemented for KSP, nanoCoP-M, and MleanCoP (hence their
integration into the TPTP World).
Thankfully this has been quite easy, and implemented mostly in {\tt sed}.

There have been several successful ATP system developments that translate/embed non-classical 
logic into a classical logic, and apply a classical logic ATP 
system~\cite{HV06,SH07,SS11,GSB17,EA+23}.
In the TPTP World the ATFLET logic embedding tool~\cite{Ste22} can be used to do a shallow 
semantic embedding of NXF/NHF problems into TFF/THF (except NHF problems can't be embedded in 
TFF)~\cite{BP13,BR13,GSB17,GS18}.
By default ATFLET produces monomorphic embeddings (TF0/TH0), but it can optionally produce 
polymorphic embeddings (TF1/TH1), which can be significantly shorter if the input problem 
contains many user types.
Currently ATFLET supports a range of modal logics, a range of first-order quantified hybrid 
logics~\cite{AtC07}, public announcement logic~\cite{vvK07,Pac13}, and two different dyadic 
deontic logics~\cite{CJ13,Aqv02}.
ATFLET can be used as a preprocessor to any TPTP-compliant TFF/THF ATP system, to form an 
ATP system for NXF/NHF.
This is the approach taken by Leo-III.

%--------------------------------------------------------------------------------------------------
\subsection{The TSTP Solution Library}
\label{TSTPSolution Library}

The TSTP solution library~\cite{Sut10} is a library of ATP systems’ solutions to TPTP problems, 
built by running all the ATP systems in the TPTP World on all the problems in the TPTP problem
library.
A major use of the TSTP is by ATP system developers, who examine solutions to understand how 
problems can be solved, leading to improvements in their own systems. 
The TSTP also provides the performance data used to compute the TPTP problems’ difficulty ratings.

%--------------------------------------------------------------------------------------------------
\subsubsection{Derivations}
\label{Derivations}

A derivation written in the TPTP language is a list of annotated formulae~\cite{SS+06}.
The leaves of a derivation typically have the role \smalltt{axiom} or \smalltt{conjecture}, 
and the inferred formulae typically have the role \smalltt{plain} or \smalltt{negated\_conjecture}.
The source record of each annotated formula is either a \smalltt{file} record for leaves, or 
an \smalltt{inference} record for inferred formulae.
A \smalltt{file} record contains the problem file name and the corresponding annotated formulae 
name in the problem file.
An \smalltt{inference} record contains the inference rule name, a list of useful inference 
information, and a list of the inference parents.
Common types of useful inference information are a \smalltt{status} record recording the semantic 
relationship of the inferred formula to its parents as an SZS ontology value~\cite{Sut08-KEAPPA},
special information about recognized types of complex inference rules, e.g., Skolemization,
and details of new symbols introduced in the inference.
Inference parents can be annotated formulae names, or nested \smalltt{inference} records.
% The SZS values are used in GDV's approach to verification (see Section~\ref{Verifiers}).
Figure~\ref{THFDerivationExample} shows some parts of Leo-III's proof for the problem
in Figure~\ref{THFExample}.

\begin{figure}[h!]
\small
\setstretch{0.9}
\begin{verbatim}
%------------------------------------------------------------------------------
thf(sk1_type,type, sk1: $i > $i > $o ).
thf(sk2_type,type, sk2: ( $i > $o ) > $i ).

thf(1,conjecture,
    ~ ? [A: $i > $i > $o] :
      ! [B: $i > $o] :
      ? [C: $i] : ( ( A @ C ) = B ),
    file('SET557^1.p',surjectiveCantorThm) ).

thf(2,negated_conjecture,
    ~ ~ ? [A: $i > $i > $o] :
        ! [B: $i > $o] :
        ? [C: $i] : ( ( A @ C ) = B ),
    inference(neg_conjecture,[status(cth)],[1]) ).

    ... some formulae omitted

thf(32,plain,
    ~ ( sk1
      @ ( sk2 @ ^ [A: $i] : ~ ( sk1 @ A @ A ) )
      @ ( sk2 @ ^ [A: $i] : ~ ( sk1 @ A @ A ) ) ),
    inference(pre_uni,[status(thm)],[18]) ).

thf(381,plain,
    $false,
    inference(rewrite,[status(thm)],[272,32]) ).
%------------------------------------------------------------------------------
\end{verbatim}
\caption{Parts of a proof for the NHF problem in Figure~\ref{THFExample}, in THF.}
\label{THFDerivationExample}
\end{figure}

The TPTP format for derivations can be used for writing derivations in non-classical logics, 
potentially making use of the new NTF connectives.
Of the non-classical ATP systems that have been integrated into the TPTP World (see 
Section~\ref{ATPSystems}), only KSP outputs proofs in non-classical logic using non-classical
connectives, but the proofs are not in TPTP format.
nanoCoP-M and MleanCoP output proofs in their own format.
Leo-III outputs proofs in TPTP format, but the proofs are in the TFF/THF language of the embedded 
problems (see Section~\ref{ATPSystems}).

Figure~\ref{NX0Proof} shows parts of a proof found by Leo-III for the problem in 
Figure~\ref{NX0Example}.
% (see~\cite{Ste18,SB21} for details of the proof search and construction process of Leo-III).
In this context, Leo-III makes use of the first three of four new defined symbols that were 
introduced for representing Kripke interpretations (see Section~\ref{Interpretations}).
These symbols are:
\begin{packed_itemize}
\item The defined type {\tt \$world} for worlds.
      Different constants of type {\tt \$world} are known to be unequal.
\item The defined predicate {\tt \$accessible\_world} of type {\tt (\$world\,*\,\$world)\,>\,\$o}
      to specify accessibility between worlds.
\item The defined constant {\tt \$local\_world} of type {\tt \$world} that identifies the world 
      in which the conjecture is proved. 
\item The defined predicate {\tt \$in\_world} of type {\tt (\$world\,*\,\$o)\,>\,\$o} to provide
      worldly scope for a formula.
\end{packed_itemize}
Leo-III's use of semantic embedding means that the meta-logical concepts of modal logic are lifted 
to the TFF/THF object logic, and hence modal logic proofs by Leo-III intrinsically encode modal 
logic semantics. 
This is made transparent by use of the defined symbols.
Modal logic proofs in a modal logic calculus, such as produced by KSP, do not need these symbols.

\begin{figure}[h!]
\small
\setstretch{0.85}
\begin{verbatim}
%------------------------------------------------------------------------------
thf(sk1_type,type, sk1: person > $world > $world ).

thf(m_reflexive,axiom,
    ! [A: $world] : ( $accessible_world @ A @ A ),
    file('LeoWorkers.p',mrel_reflexive) ).

thf(alex_works_on_leo,axiom,
    work_hard @ $local_world @ alex @ leo,
    file('LeoWorkers.p',alex_works_on_leo_here) ).

thf(work_possible_rich,axiom,
    ! [A: $world,B: person] :
      ( ? [C: product] : ( work_hard @ A @ B @ C )
     => ? [C: $world] :
          ( ( $accessible_world @ A @ C ) & ( gets_rich @ C @ B ) ) ),
    file('LeoWorkers.p',work_hard_to_get_rich) ).

thf(cnf_work_possible_rich_1,plain,
    ! [C: product,B: person,A: $world] :
      ( ~ ( work_hard @ A @ B @ C ) | ( gets_rich @ ( sk1 @ B @ A ) @ B ) ),
    inference(cnf,[status(esa)],[work_possible_rich]) ).

    ... formulae omitted here 

thf(prove_inequity,conjecture,
    ? [A: person,B: $world] :
      ( ( $accessible_world @ $local_world @ B )
      & ( gets_rich @ B @ A ) & ~ ( gets_rich @ B @ ( advisor_of @ A ) ) ),
    file('LeoWorkers.p',someone_gets_rich_but_not_advisor) ).

thf(neg_prove_inequity,negated_conjecture,
    ~ ? [A: person,B: $world] :
        ( ( $accessible_world @ $local_world @ B )
        & ( gets_rich @ B @ A ) & ~ ( gets_rich @ B @ ( advisor_of @ A ) ) ),
    inference(neg_conjecture,[status(cth)],[prove_inequity]) ).

    ... formulae omitted here 

thf(732,plain,
    ! [B: $world,A: $world] :
      ( ~ ( gets_rich @ B @ alex )
      | ( ( $accessible_world @ A @ A )
       != ( $accessible_world @ ( sk1 @ alex @ $local_world ) @ B ) ) ),
    inference(paramod_ordered,[status(thm)],[m_reflexive,591]) ).

thf(733,plain,
    ~ ( gets_rich @ ( sk1 @ alex @ $local_world ) @ alex ),
    inference(pattern_uni,[status(thm)],[732]) ).

thf(772,plain,
    $false,
    inference(rewrite,[status(thm)],[733,336]) ).
%------------------------------------------------------------------------------
\end{verbatim}
\caption{Parts of a proof for the NX0 problem in Figure~\ref{NX0Example}, in THF.}
\label{NX0Proof}
\end{figure}

%--------------------------------------------------------------------------------------------------
\subsubsection{Interpretations}
\label{Interpretations}

The TPTP format for interpretations~\cite{SS+23-LPAR,SSF24} can express both Tarskian~\cite{TV56}
and Kripke~\cite{Kri63} interpretations.
Figure~\ref{TFFTarskianExample} shows parts of a Tarskian interpretation for the first-order 
aspects of Figure~\ref{NX0Example}.
Note how function and predicate symbols are interpreted by applying them directly to domain
elements, \emph{{\`a} la}~\cite[\S5.3.4]{Gal15}, rather than introducing a new interpreting 
function for each function/predicate symbol.
% \emph{{\`a} la}~\cite[\S5.3.2]{Gal15}. 
The direct application requires using \emph{type-promotion} bijections to keep formulae 
well-typed, e.g., {\tt d2P} in Figure~\ref{TFFTarskianExample}.
A TPTP format Kripke interpretation provides information about the worlds and their accessibility,
and a Tarskian interpretation for each world.
Figure~\ref{TXFKripkeExample} shows parts of a Kripke model for the axioms in 
Figure~\ref{NX0Example}.

\begin{figure}[h!]
\small
\setstretch{0.9}
\begin{verbatim}
%------------------------------------------------------------------------------
%----Problem types omitted
%----Domain types
tff(d_person_type,type,  d_person: $tType ).
tff(d_product_type,type, d_product: $tType ).
%----Domain elements
tff(dp_1_decl,type,      dp_1: d_person ).
tff(dp_2_decl,type,      dp_2: d_person ).
tff(dr_1_decl,type,      dr_1: d_product ).
tff(dr_2_decl,type,      dr_2: d_product ).
%----Type promotions
tff(d2p_decl,type,       d2p: d_person > person ).
tff(d2r_decl,type,       d2r: d_product > product ).

tff(leo_workers_domains,interpretation-domain,
    ( ! [P: person] : ? [DP: d_person] : ( P = d2p(DP) )
    & ! [DP: d_person] : ( ( DP = dp_1 ) | ( DP = dp_2 ) )
    & ( dp_1 != dp_2 )
    & ! [DP1: d_person,DP2: d_person] : 
        ( ( d2p(DP1) = d2p(DP2) ) => ( DP1 = DP2 ) )
    & ! [R: product] : ? [DR: d_product] : ( R = d2r(DR) )
    & ! [DR: d_product] : ( DR = dr_1 )
    & ! [DR1: d_product,DR2: d_product] : 
        ( ( d2r(DR1) = d2r(DR2) ) => ( DR1 = DR2 ) ) ) ).

tff(leo_workers_mappings,interpretation-mapping,
    ( leo = d2r(dr_1)
    & ( advisor_of(d2p(dp_1)) = d2p(dp_2) )
    & ( advisor_of(d2p(dp_2)) = d2p(dp_1) )
    & ~ gets_rich(dp_1)
    & ~ gets_rich(dp_2)
    & work_hard(d2p(dp_1),d2r(dr_1))
    & work_hard(d2p(dp_2),d2r(dr_1))
    & ~ work_hard(d2p(dp_1),d2r(dr_2))
    & ~ work_hard(d2p(dp_2),d2r(dr_2)) ) ).
%------------------------------------------------------------------------------
\end{verbatim}
\caption{A Tarskian interpretation for the local world of Figure~\ref{NX0Example}.}
\label{TFFTarskianExample}
\end{figure}

\begin{figure}[h!]
\small
\setstretch{0.9}
\begin{verbatim}
%------------------------------------------------------------------------------
%----Problem and Tarskian types omitted
%----Worlds
tff(w1_decl,type,w1: $world).
tff(w2_decl,type,w2: $world).

tff(leo_workers,interpretation-worlds,
    ( ! [W: $world] : ( W = w1 | W = w2 )
    & $local_world = w1
    & $accessible_world(w1,w1) 
    & $accessible_world(w2,w2)
    & $accessible_world(w1,w2)
    & $accessible_world(w2,w2) ) ).

tff(leo_workers,interpretation,
    ( $in_world(w1,
        ( ! [P: person] : ? [DP: d_person] : ( P = d2p(DP) )
    ... copied from the Tarskian interpretation
        & work_hard(d2p(dp_2),d2r(dr_1))
        & ~ work_hard(d2p(dp_1),d2r(dr_2))
        & ~ work_hard(d2p(dp_2),d2r(dr_2)) ) )
    & $in_world(w2,
    ... etc.
        & ~ work_hard(d2p(dp_2),d2r(dr_1))
        & ~ work_hard(d2p(dp_1),d2r(dr_2))
        & ~ work_hard(d2p(dp_2),d2r(dr_2)) ) )
%------------------------------------------------------------------------------
\end{verbatim}
\caption{Parts of a Kripke interpretation for the axioms in TXF.}
\label{TXFKripkeExample}
\end{figure}

%--------------------------------------------------------------------------------------------------
\subsubsection{NTF Solutions in the TSTP}
\label{NTFSolutions}

The TSTP solution library includes the results of running the ATP systems KSP~0.1.7, nanoCoP-M~2.0,
MleanCoP~1.3, and Leo-III~1.7.18 on the 215 NTF problems in the TPTP problem library v9.1.0.
Additionally, E~3.2.5 and Vampire~4.10 were tested on the NTF problems embedded into TF0 and TH0 
using ATFLET (see Section~\ref{ATPSystems}).
Recall that KSP is limited to the 52 propositional problems, nanoCoP-M and MleanCoP are limited
to the 25 first-order problems with only local terms, local axioms, non-decreasing domains, 
and {\bf D}/{\bf M}/{\bf S4}/{\bf S5} modalities, and the 21 NH0 problems cannot be embedded 
into TFF.
The results are shown in Table~\ref{Results}, for all the propositional problems and all the
non-propositional theorems, which is 179 of the 215 problems in the TPTP.
KSP is the most successful for propositional problems, particularly for showing that a
problem is CounterSatisfiable ({\tt CSA}).
For the non-propositional problems there is very little difference between the systems that
run on the ATFLET embeddings, which suggests that the embedding is a dominant influence.
There is only a slight advantage to embedding NX0 problems into TF0 rather than TH0.

\begin{table}[htb]
\setlength{\tabcolsep}{4pt}
\begin{tabular}{lr|rrrrrrrr}
Type                       &  \# & KSP & n'CoP & M'CoP & Leo & E-TF0 & E-TH0 & V-TF0 & V-TH0 \\
\hline
NX0 PRP THM                &  27 &   7 &     2 &     4 &   6 &     7 &     5 &   {\bf 10} &     6 \\
NX0 PRP CSA                &  25 & {\bf 14} &    10 &    10 &   0 &     5 &     0 &     8 &     0 \\
NX0 THM                    & 107 &   - &     5 &     5 & {\bf 84} &   {\bf 84} &    82 &   {\bf 84} &    82 \\
\hspace*{0.2cm}NX0 THM NEQ &  96 &   - &     5 &     5 &  75 &    75 &    75 &    75 &    75 \\
\hspace*{0.2cm}NX0 THM EQU &  11 &   - &     0 &     0 &  {\bf 9} &    {\bf 9} &     7 &    {\bf 9} &     7 \\
NH0 THM                    &  20 &   - &     0 &     0 &  19 &     - &    19 &     - &    19 \\
\hspace*{0.2cm}NH0 THM NEQ &   9 &   - &     0 &     0 &   8 &     - &     8 &     - &     8 \\
\hspace*{0.2cm}NH0 THM EQU &  11 &   - &     0 &     0 &  11 &     - &    11 &     - &    11 \\
\hline
Total                      & 179 &  21 &    17 &    19 & 109 &    96 &   106 &   102 &   107 \\
\end{tabular}
\caption{Results from testing ATP systems on non-classical logic problems in TPTP v9.1.0.}
\label{Results}
\end{table}

%--------------------------------------------------------------------------------------------------
\subsection{Tool Support}
\label{Tools}

TPTP World software support for non-classical logics has been developed, and continues to be 
developed.
All the software is freely available from GitHub\footnote{%
\href{https://github.com/TPTPWorld}{\tt github.com/TPTPWorld}}.
As a convenient alternative to downloading and installing TPTP World tools locally, there are
three online interfaces that provide access to ATP systems and tools \cite{Sut00-CADE-17,Sut07-CSR}:
\begin{itemize}
\item SystemB4TPTP\footnote{%
      \href{https://tptp.org/cgi-bin/SystemB4TPTP}{\tt tptp.org/cgi-bin/SystemB4TPTP}}
      for preparing formulae (often problems) for subsequent submission to ATP systems and other
      tools.
      SystemB4TPTP includes the ATFLET tool and syntax translators (see Section~\ref{ATPSystems}).
\item SystemOnTPTP\footnote{%
      \href{https://tptp.org/cgi-bin/SystemOnTPTP}{\tt tptp.org/cgi-bin/SystemOnTPTP}}
      for submitting formulae to ATP systems.
      SystemOnTPTP includes the ATP systems for NTF (see Section~\ref{ATPSystems}).
\item SystemOnTSTP\footnote{%
      \href{https://tptp.org/cgi-bin/SystemOnTSTP}{\tt tptp.org/cgi-bin/SystemOnTSTP}}
      for processing solutions output by ATP systems and other tools.
\end{itemize}

The TPTP2T tool for listing problems and solutions with specified syntactic and semantic 
characteristics is available in a separate interface\footnote{%
\href{https://tptp.org/cgi-bin/TPTP2T}{\tt tptp.org/cgi-bin/TPTP2T}}. 
It could be used, e.g., to list NX0 problems that are theorems, contain equality, and are easy 
to prove.

%--------------------------------------------------------------------------------------------------
\subsubsection{Parsers and Printers}
\label{Parsers}

The TPTP4X utility and the BNF-based suite of parsers~\cite{VS06} can parse NXF and NHF formulae.
TPTP4X parses problems and solutions, can apply various transformations, and pretty-prints the 
formulae.
The BNF-based parsers offer stricter parsing than TPTP4X, and can present parse trees in
various forms.

In addition to the TPTP World's own tools, a suite of tools that can parse and manipulate 
NXF and NHF formulae is available in the Leo-III framework~\cite{SB21}.
The {\tt tptp-utils} tool~\cite{Ste22-TU} can read formulae in all the TPTP languages, including 
NXF and NHF.
It does syntax checking, translations, generation of parse trees, basic linting, and
pretty-printing.
For NXF and NHF in particular, it can sanity check logic specifications for modal logics.
% It comes with a complete definition of abstract syntax trees for the internal representation.
Its underlying parser written in Scala is available as the stand-alone parsing library 
{\tt scala-tptp-parser}~\cite{Ste21}.

%--------------------------------------------------------------------------------------------------
\subsubsection{Verifiers and Viewers}
\label{Verifiers}

The GDV derivation verifier~\cite{Sut06,SBB25} can verify proofs in TPTP format.
GDV is available as a standalone tool, and also in SystemOnTSTP.
GDV does:
\begin{packed_itemize}
\item Structural verification, e.g., checking that the derivation is acyclic.
\item Origin verification, e.g., checking that the leaves of the derivation are (derivable) from 
      the problem formulae.
\item Completeness verification, e.g., checking that the root of a refutation is {\em false}.
\item Inference verification using a trusted ATP system, e.g., checking that an inferred 
      formula is a theorem of its parents.
\end{packed_itemize}

The AGMV model verifier can verify Kripke models in TPTP format~\cite{SS+23-LPAR}.
AGMV is available as a standalone tool, and also in SystemOnTSTP.
AGMV does:
\begin{packed_itemize}
\item Syntax and type checking. 
\item Model consistency checking using a trusted model finder, to ensure that the interpretation 
      formulae are satisfiable,
\item Model correctness checking using a trusted theorem prover, to confirm that the problem 
      formulae are theorems of the interpretation formulae.
\end{packed_itemize}
% AGMV has been tested with examples created by hand.

\begin{figure}[htbp]
\centering
\includegraphics[width=0.80\textwidth]{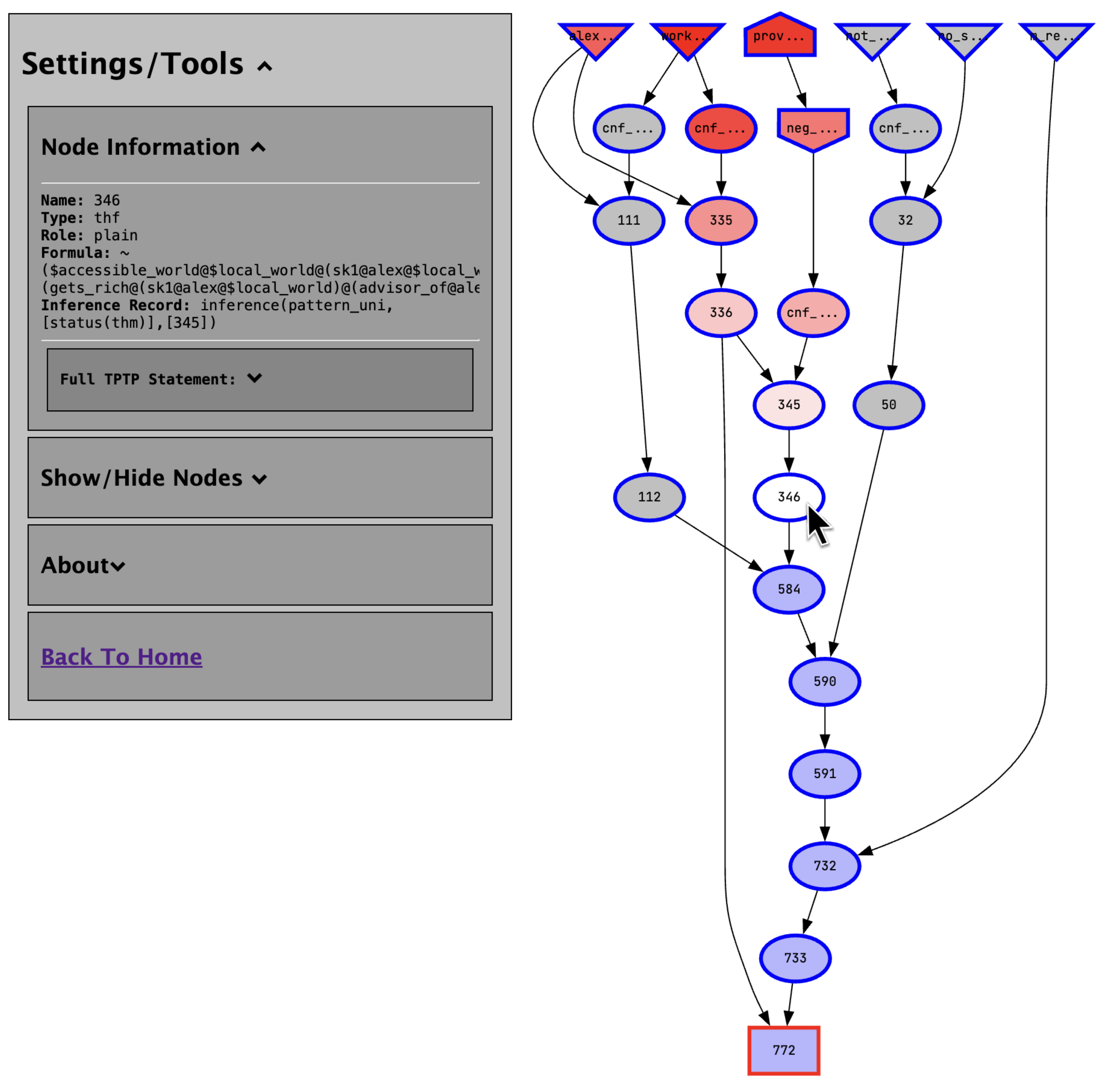}
\caption{IDV view of the proof in Figure~\ref{NX0Proof}.}
\label{NX0ProofIDV}
\end{figure}

The IDV Interactive Derivation Viewer~\cite{TPS07} is able to display derivations in TPTP format.
Figure~\ref{NX0ProofIDV} shows the NX0 proof in Figure~\ref{NX0Proof}.
The pointer is hovering over the node {\tt 346}, whose formula is shown in the lefthand panel.
The red and blue coloring shows the ancestors and descendants of the node in the derivation.
IDV is available in SystemOnTSTP.

\begin{figure}[htbp]
\centering
\includegraphics[width=0.80\textwidth]{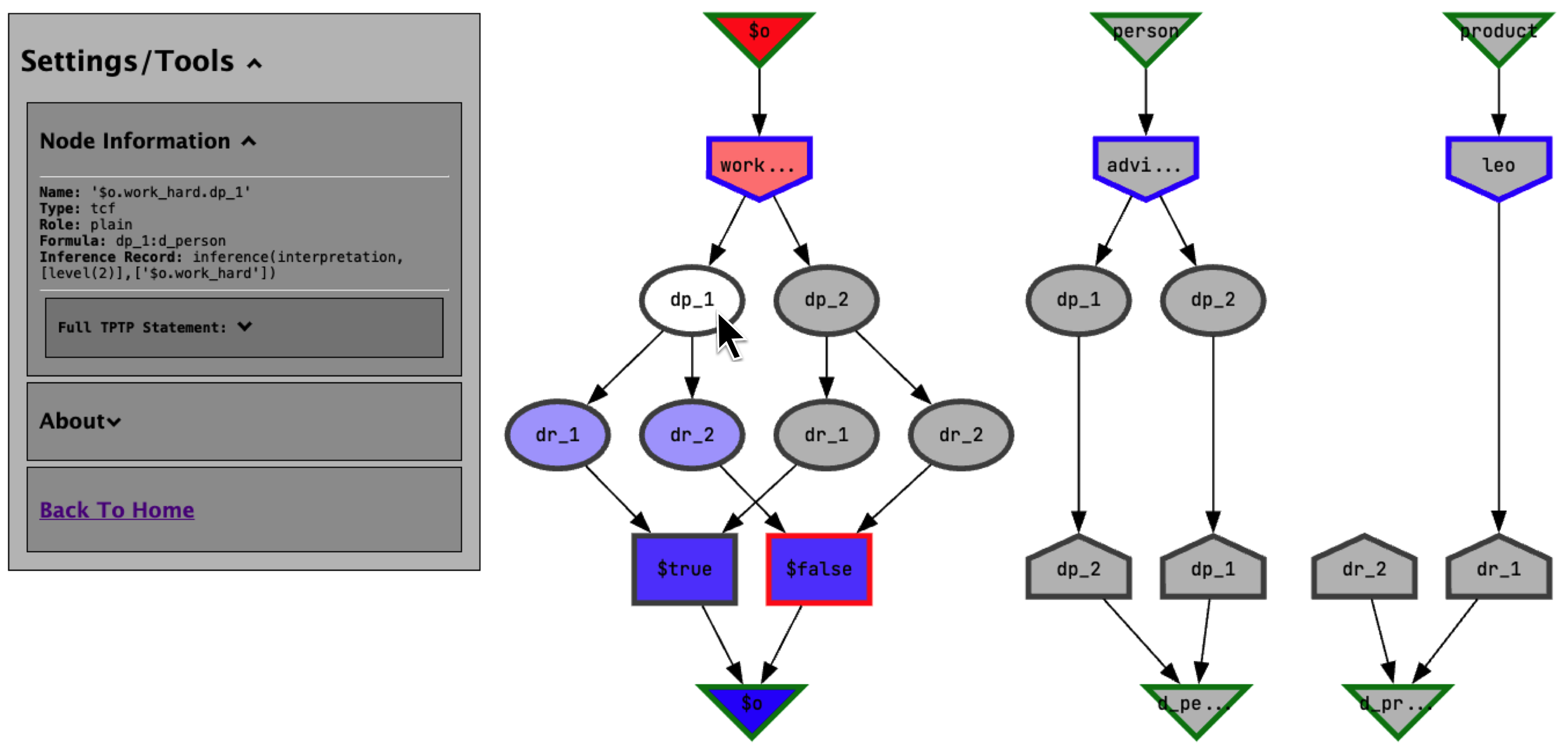}
\caption{IIV view of the Tarskian interpretation in Figure~\ref{TFFTarskianExample},
         which is the Tarskian interpretation in world {\tt w1} in Figure~\ref{TXFKripkeExample}.}
\label{NX0InterpretationW1}
\end{figure}

The IIV interactive interpretation viewer~\cite{SS+23-LPAR,MS23-Poster} is able to display the
TPTP format Tarskian interpretation of a world in a Kripke model.
Figure~\ref{NX0InterpretationW1} shows the interpretation of world {\tt w1} in 
Figure~\ref{TXFKripkeExample}.
The pointer is hovering over the node {\tt dp\_1}.
The red {\tt \$o} ancestor indicates that {\tt work\_hard} has a boolean result type.
The blue {\tt \$false} descendant indicates that the interpretation of {\tt work\_hard} for the
domain elements {\tt dp\_1} and {\tt dr\_2} is {\em false}, and the blue {\tt \$o} descendant 
indicates that the domain type is boolean.
IIV is available in SystemOnTSTP.
A wrapper to view the worlds and accessibility relationships of a Kripke model needs to be 
developed and linked to IIV to view a chosen world's Tarskian interpretation.

%--------------------------------------------------------------------------------------------------
\section{Summary and Discussion}
\label{Conclusion}

This paper provides a comprehensive overview of the support for non-classical logics
in the TPTP World, including languages, problems, solutions, and infrastructure. 
The general non-classical format is exemplarily instantiated with quantified normal multi-modal 
logic.
The TPTP World's non-classical NTF language is a conservative syntactic extension of the 
classical TXF and THF languages. 
The extension is deliberately minimal (only one new type of connective), general (non-classical 
connectives can have arbitrarily many arguments and meta-logical parameters), flexible (users can 
define their own logics and connectives), and consistent with existing TPTP principles (e.g., 
Prolog-compatibility). 
Logic specifications are introduced to provide relevant information about the non-classical 
logic in which a problem is formulated.
Modal logic problems have been part of the TPTP problem library since release v9.0.0.

It is apparent that the general-purpose NTF language supports many non-classical logics, and thus 
is not as concise as special-purpose formats for specific logics (although the fact that {\tt sed} 
can be used to translate NTF to the specialised KSP, nanoCoP-M, and MleanCoP formats shows that 
NTF is quite compatible with those special-purpose formats (see Section~\ref{ATPSystems})).
The cost of generality is outweighed by the advantages gained:
(i)~The distinction between meta-logical parameters and object-logic arguments of non-classical 
connectives is clear. 
In other languages meta-logical parameters are often encoded by increasing the arity of the 
connectives, and treating the parameters as additional arguments that might require type
declarations. 
This is avoided in NTF, eliminating potential typing issues.
(ii)~Problem files document the logic in which the problem is formulated, using a logic 
specification. 
This increases clarity and reproducibility, as the information about the theoremhood of the 
conjecture or (un)satisfiability of the formulae does not depend on additional external 
information. 
Other languages often leave out this information, and ATP systems have to be given command-line 
parameters to specify what logic is to be used.

Further work includes:
\begin{packed_itemize}
\item Collecting more quantified modal logic problems for the TPTP problem library.
\item Standardizing further popular non-classical logics in NTF
      (obvious candidates include linear logics and intuitionistic logics).
\item Working with ATP system developers to upgrade their systems to natively read problems 
      written in NXF and NHF, and to produce proofs and models in TPTP format in the same 
      non-classical logic as used for the problem.
\item Producing a complete interactive viewer for Kripke interpretations written in TPTP-format.
\item Extending the logic specification format to allow for combinations of logics.
\item A non-classical logic division of CASC.
\end{packed_itemize}

\section*{Acknowledgements}
The authors sincerely thank Christoph Benzmüller for stimulating discussions and helpful comments.

\section*{Author Contribution Statement}
Both authors contributed equally.

\section*{Funding Declaration}
The authors did not receive funding for this research.

%--------------------------------------------------------------------------------------------------
\bibliography{Bibliography}
%--------------------------------------------------------------------------------------------------
\end{document}